%% file: main.tex
\pdfoutput=1
\documentclass[10pt,journal,compsoc]{IEEEtran}
\usepackage[hidelinks]{hyperref}
\usepackage{color}
\usepackage{graphicx}
\usepackage{comment}
\usepackage{xspace}
\usepackage{todonotes}
\usepackage{amsmath}
\usepackage{enumitem}
\usepackage{framed}
\usepackage{xcolor}
\usepackage{flushend}

\newif\ifFinal
\Finaltrue
\ifFinal
\newcommand{\saurabh}[1]{}
\newcommand{\fred}[1]{}
\newcommand{\todoinline}[1]{}  
\newcommand{\milind}[1]{} 
\newcommand{\saurabhtodo}[1]{}
\else
\newcommand{\saurabh}[1]{\textcolor{blue}{SB: #1}}
\newcommand{\fred}[1]{\textcolor{green}{\bf FD: #1}\xspace}
\newcommand{\milind}[1]{\textcolor{purple}{\bf MK: #1}\xspace}
\newcommand{\todoinline}[1]{\todo[inline,color=green!40]{ToDo: [#1]}}
\newcommand{\saurabhtodo}[1]{\todo[inline,color=green!40]{saurabh: #1}}
\fi

\newcommand{\eg}{{\em e.g.},\xspace}
\newcommand{\ie}{{\em i.e.},\xspace}

\begin{document}
\title{Vision Paper: \\ Grand Challenges in Resilience: Autonomous System Resilience through Design and Runtime Measures}
\author{Saurabh Bagchi$^{\psi,\alpha}$, 
Vaneet Aggarwal$^\alpha$, 
Somali Chaterji$^\alpha$, 
Fred Douglis$^\beta$, 
Aly El Gamal$^\alpha$, 
Jiawei Han$^\gamma$, 
Brian J. Henz$^\delta$, 
Hank Hoffmann$^\epsilon$, 
Suman Jana$^\zeta$, 
Milind Kulkarni$^\alpha$, 
Felix Xiaozhu Lin$^\alpha$, 
Karen Marais$^\alpha$, 
Prateek Mittal$^\eta$, 
Shaoshuai Mou$^\alpha$, 
Xiaokang Qiu$^\alpha$, 
Gesualdo Scutari$^\alpha$ \\
\vspace{3em}
{\small $\psi$: Corresponding author. All other authors are listed alphabetically. \\
$\alpha$: Purdue University, $\beta$: Perspecta Labs, $\gamma$: University of Illinois at Urbana-Champaign, $\delta$: Army Research Lab, $\epsilon$: University of Chicago, $\zeta$: Columbia University, $\eta$: Princeton University}
}
% \date{}
\maketitle

%\textbf{Limit: While we have not finalized which magazine will be the home for this article, in terms of length, we are targeting the length of a CACM ``Contributed Article": 4000 words, about 12 pages in this format.}

\input{abstract}

\input{introduction}

\input{autonomous-systems}

\input{resilient-design}

\input{execution-platforms}

\input{road-ahead}

\section*{Acknowledgments}

This work is supported in part through NSF Award for the Grand Challenges in Resilience Workshop at Purdue March 19--21, 2019 through award number CNS-1845192. DEDUCE was sponsored by DARPA under the EdgeCT program HR0011-15-C-0098. Any opinions, findings, and conclusions or recommendations expressed in this material are those of the authors and do not necessarily reflect the views of the sponsors.

\begin{small}
\bibliographystyle{plain}
\bibliography{sbagchi,Ref_Res,schaterji,refhan,refqiu,refaly,fdouglis}
\end{small}

\end{document}

%% file: abstract.tex
\section{Abstract}
\label{sec:abstract}

% \textcolor{red}{SC: I removed the workshop-report style and made the format more like a vision paper, may help with the CACM-style submission. Will need to add a few more motivating lines for cyber resilience upfront.}
% \saurabh{The intro part is needed for the workshop report. This part (till "In this article, we put forward ..." is to be removed for the OJCS submission.}
% SB: Modified for OJCS submission.

% A set of about 80 researchers, practitioners, and federal agency program managers participated in the NSF-sponsored Grand Challenges in Resilience Workshop held on Purdue campus on March 19--21, 2019~\cite{crisp-workshop-19}. The workshop was divided into three themes: resilience in cyber, cyber-physical, and socio-technical systems. About 30 attendees in all participated in the discussions of cyber resilience. 
% This article brings out the substantive challenges to building resilient autonomous systems and solution approaches that were identified in the cyber resilience theme. 
% We focused our discussion around three broad topics within the theme. \\
% Our article develops its ideas using these topics as anchor points and extrapolates from them to the broader theme of cyber resilience. It develops them under the themes of: \\
In this article, we put forward the substantial challenges in cyber resilience in the domain of autonomous systems and outline foundational solutions to address these challenges. These solutions fall into two broad themes:  \textit{resilience-by-design} and \textit{resilience-by-reaction}.
We use several application drivers from  autonomous systems to motivate the challenges in cyber resilience and to demonstrate the benefit of the solutions. We focus on some autonomous systems in the near horizon (autonomous ground and aerial vehicles) and also a little more distant (autonomous rescue and relief). \\
For {\em resilience-by-design}, we focus on design methods in software that are needed for our cyber systems to be resilient.
In contrast, for {\em resilience-by-reaction}, we discuss how to make systems resilient by responding, reconfiguring, or recovering at runtime when failures happen. We also discuss the notion of adaptive execution to improve resilience, execution transparently and adaptively among available execution platforms (mobile/embedded, edge, and cloud). 
For each of the two themes, we survey the current state, and the desired state and ways to get there. We conclude the paper by looking at the research challenges we will have to solve in the short and the mid-term to make the vision of resilient autonomous systems a reality. This article came out of discussions that started at the NSF-sponsored Grand Challenges in Resilience Workshop held at Purdue in 2019 with the co-authors contributing to going into the depth of the issues and then this article. 

\begin{comment}

\begin{enumerate}
    \item {\em Autonomous systems}: The application drivers. We focus on some autonomous systems in the near horizon (autonomous vehicles, both cars and aerial vehicles) and some a little more distant (personalized smart city services). 
    
    \item {\em Resilience by design}: The suite of software techniques that can be used to make systems resilient prior to deployment. 
    
% I'm not sure what "flitting" means here.  Do you mean "fitting"?  Either way it seems odd - FD
% SB (5/25/19): Reworded and resolved.
    \item {\em Mobile, edge, and cloud platforms}: The universe of platforms on which the applications will execute in a resilient manner. There is a desire to have flexible and {\em not} pre-programmed migration of the relevant parts of the application among these three execution platforms. The migration can happen due to the need to meet performance guarantees or to handle failures. 
    
\end{enumerate}

\end{comment}

%% file: introduction.tex
\section{Introduction}
\label{sec:introduction}

We lay out our vision for resilience in autonomous systems and our view of the short-term and mid-term research challenges to realize the vision. Our view of resilience has two primary aspects.

\begin{enumerate}
    \item {\em Resilience by design}: This is the aspect that designs and develops cyber systems so that they are resilient to a large set of quantifiable {\em perturbations}.
    
    \item {\em Resilience by reaction}: This is the aspect that works at runtime when perturbations are incident on the cyber system and imbues the systems with the ability to ``bounce back" quickly after a failure triggered by a perturbation. 
\end{enumerate}

Note, of course, that these two aspects of resilience are intertwined: systems can be {\em designed} so that they incorporate resilience by {\em reaction}.

We also make specific the notion of {\em \bf perturbations} that we want to deal with. These take three forms: (i) {\em natural failures of hardware or software} (due to bugs, aging, misconfigurations, resource contentions in shared environments, downtime due to planned upgrades, etc.), (ii) {\em maliciously induced failures} or security attacks (from outside the system), and (iii) {\em unexpected inputs} (our target class of autonomous systems will have to deal with the physical environment and will interface with humans, which will produce unpredictable data to which the system will need to adapt). 
% \fred{I think one thing that is missing here is perturbations due to performance issues.  You can have \textit{failures} as well as attacks or unexpected inputs, but you can also have performance issues that are not really failures but have similar consequences.}
% \saurabh{Added soft failures (performance failures) in addition to hard failures and attacks.}
% \fred{But I still think it is confusing.  You list natural failures, malicious failures, and unexpected inputs, then you say the outcome can be a soft failure.  I am saying that resource contention can \textit{simulate} the effect of a natural or malicious failure, but not be a failure itself.  Perhaps "due to bugs, aging, misconfiguration" should also list "resource contention" as a cause of natural failures? }
% \milind{Along the lines of what Fred mentions, can we also add something about ``planned'' migration of hardware/software in response to things like resource contention, increased demand, etc.}
% \saurabh{Taken all feedback about definition of perturbations into account. Take a look.}
The outcome of a perturbation that is not handled can be a hard failure (crash or hang) or a soft failure (\ie missing a deadline for a latency-sensitive application). 

We will first introduce as application drivers two autonomous application scenarios where perturbations need to be handled. We will then discuss the resilience by design aspect and then the resilience by reaction aspect. For each, we will lay out the vision for the end state in 10 years. Then we will talk of the short-term and mid-term research 
% and policy 
% SB (7/29/19): Shall we include policy challenges? My personal take is not as that will make it too diffuse plus we do not have the right expertise in our group. 
challenges, side-by-side with the promising approaches being investigated today. 

\begin{figure*}
\centering 
\includegraphics[width=0.75\linewidth]{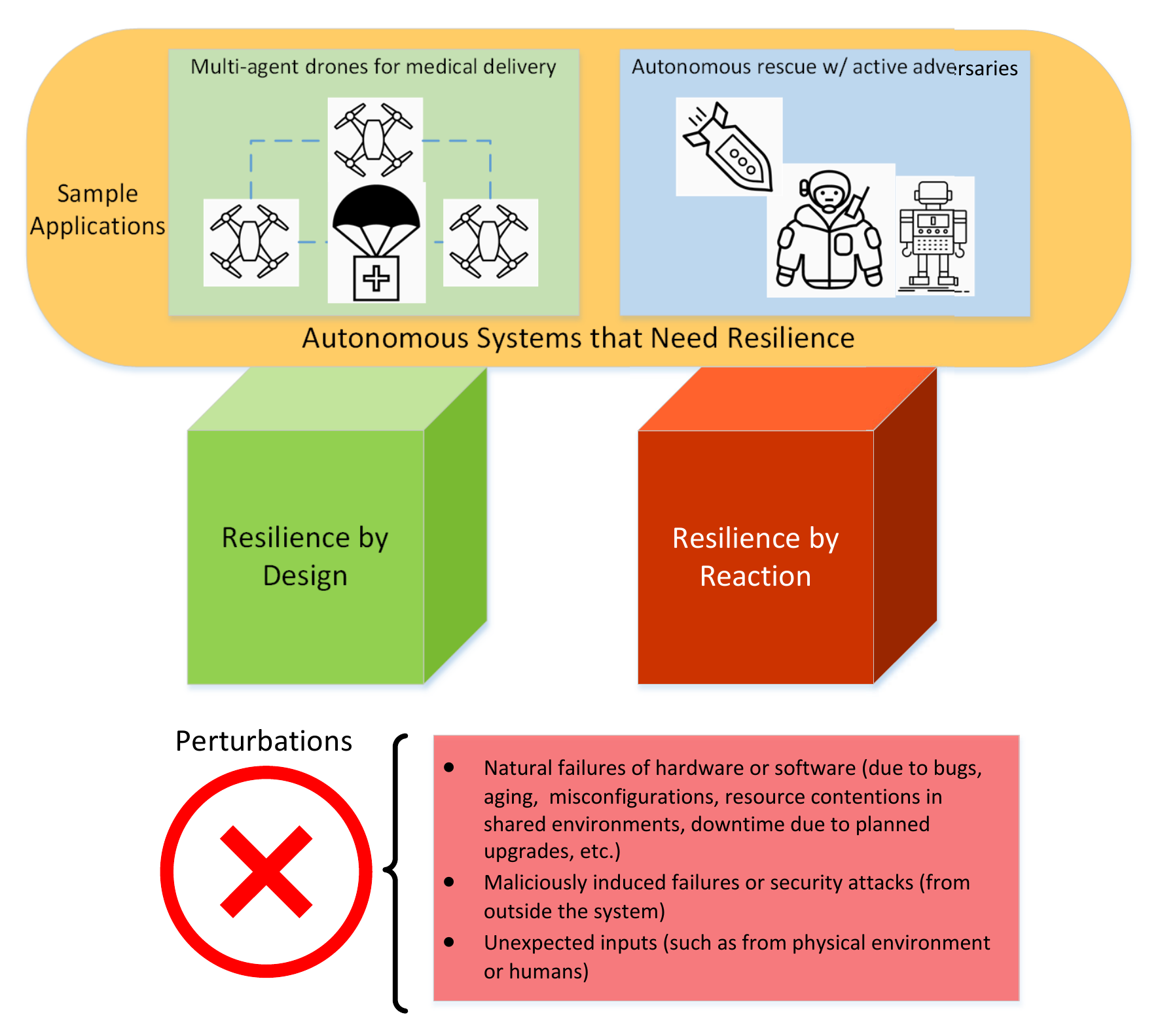}
\caption{High-level schematic of our vision for how resilience for autonomous systems can be structured. There are interplays between the two thrusts --- resilience by design (offline techniques) and resilience by reaction (online techniques). We also show the target classes of perturbations that we consider in this article.}
\label{fig:overview-top-level}
\end{figure*}

Figure~\ref{fig:overview-top-level} shows a high-level schematic for the way we envision resilience in autonomous systems together with the universe of target perturbations. 

%% file: autonomous-systems.tex
\section{Autonomous Systems as Application Drivers}
\label{sec:autonomous-systems}

\saurabh{Contributors (open to suggestions): Lead -- Saurabh Bagchi; Somali Chaterji, Karen Marais, Shaoshuai Mou, David Corman, Jennifer Weller, Brian Henz}
% Elsa Gonsiorowski

We first look at the state of resilience in two exemplar classes of autonomous systems. By autonomous we do not mean completely autonomous, rather those at varying levels of autonomy which involve some human involvement. We speculate at a desired degree of resilience against perturbations and use these as examples as broad motivation for the solution directions that we lay out in the rest of the article.  

The first class of autonomous systems is drones being used to deliver essential medical supplies in hard-to-reach areas. We include in our
purview interactions among multiple drones and among drones and the non-(Computer Science)expert humans responsible for their resilient operation. The second class of autonomous systems is rescue and relief by say an international humanitarian agency in the face of a natural or a man-made disaster. This involves ground and aerial sensors, distributed inferencing from their inputs through processing at the edge as well as at the cloud. 

\begin{comment}
Possible autonomous systems to use as running examples: 
Drones - for delivery of essential medicines, and medical supplies \\
\karen other drone uses: property surveying, natural disaster surveying, delivery of all kinds of stuff (from pizza to medicine)
Robotic surgery/telemedicine - for serving rural locations where medical facilities and qualified health practitioners are lacking  \\
\karen are we really suggesting autonomous surgery? I think with an example like that we should carefully define what we mean
\todo{Add other possible autonomous systems here. Bring in the notion of multiple autonomous agents needing to act together, also humans in the loop.}
\karen drone surveying with multiple drones; drones providing mobile internet; drones operating with manned aircraft in controlled airspace
\end{comment}

% \todoinline{I suggest shortening this following part considerably and keeping the discussion at a high-level to show what kinds of perturbations are possible, how they are poorly handled today, and what a desired end state can be. Thus we can skip the lengthy amount of prior work discussion and open problems (that will be solved by techniques that we discuss in the next two sections).}

%%%% Input from Shaoshuai Mou %%%%%%%%%%%%%%%%%%%%%%%
% \subsection*{Towards Distributed Resilience in Multi-Agent Networks 
\subsection*{Distributed Resilience in Multi-Agent Drones for Medical Deliveries}
\saurabh{Added by Shaoshuai}

%\todoinline{[Shaoshuai] Some references are missing.}
\noindent {\bf Problem and Current State}
The field of systems and control has recently been evolving from single monolithic system to teams of interconnected subsystems (or multi-agent networks). Because of the absence of centralized coordinator,  algorithms for coordination in multi-agent networks (especially large-scale autonomous swarms) must be \emph{distributed}, which achieve global objectives through only local coordination among nearby agents \cite{FJS09DC}. In order to guarantee all agents in the networks work as a cohesive whole, the concept of \emph{consensus} naturally arises, which requires all agents in the network to reach an agreement regarding a certain quantity of interest \cite{AJA03TAC,Moreau05TAC,RJR07PIEEE}. 
A specific instantiation of this general idea of multi-agent systems is a {\bf multiple drone system that is responsible for transporting essential supplies to a population affected by a natural disaster or medical supplies to a population where the surface transportation infrastructure is poor}~\cite{medical-delivery-drones, medical-delivery-drones-2}. Some characteristics relevant to our discussion are that there are multiple cooperating agents involved, there is uncertainty in the physical as well as the cyber conditions (flying conditions may be variable and the network connectivity among multiple drones may be variable, as examples of the two kinds of uncertainty), and there is also human involvement, such as, to task the drones or to refill the supplies when the drone reaches a certain height above the ground at the home base. 

Consensus is the basis of many distributed algorithms for computations \cite{SJA15TAC,XSD17TIE,XJSM19TAC}, optimization \cite{SW17TAC,TAA14TAC,Boyd11ML}, control \cite{JR04TAC,FMF13PNAS,MM10DC} in multi-agent networks. The success of consensus-based algorithms relies on the assumption that all agents in the multi-agent swarm are cooperative, that is, each agent provides its own state value to its neighbor nodes and follows a common update protocol toward network objectives \cite{RJR07PIEEE}. 
However, this particular autonomous system presents the salient challenge that it operates in an open and potentially adversarial environment, with exposure to a large and possibly unanticipated set of perturbations. On the one hand, distributed algorithms are inherently robust against individual node or link failures because of the absence of central controller; on the other hand,  the strong dependence of distributed algorithms on local coordination raises a major concern of cyber attacks to the whole network through local attacks to one or more vulnerable agents \cite{FFF13TAC}. Thus it is important to achieve \emph{resilience} in order for autonomous multi-agent swarms to be used in critical applications like the current one.

There has been significant progress made in developing robust distributed algorithms by a combination of algorithmic and system-theoretic approaches in  \cite{ZMS10TCS,FAF12TAC, NLG12ACM}. Further advancements will have to be made to handle hitherto unanticipated perturbations, to deal with the resource constraints of individual agents, to deal with time-varying characteristics such as link quality, and the possibility of multiple coordinated or uncoordinated perturbations. An entire new dimension arises due to the close interactions with humans---different patients may have different criticality requirements and these may change over short time periods, the level of cyber expertise of the human users both at the provider and at the consumer level will vary, and the time constants involved for some operations will be of human scale rather than cyber scale. Yet another dimension that needs significant research progress is scalability of these algorithms. While they have primarily been developed and evaluated under small world assumptions, they need to be re-designed or modified to operate at large scales, of the number of agents, of the distances (and latencies) involved and under the open-world assumption (new nodes can be added while the system is operational), etc.

\subsection*{Cooperative Autonomous Rescue with Active Adversary}
\saurabh{Added by Brian Henz}

\noindent Operationalizing artificial intelligence (AI) for military applications often brings to mind either offensive or defensive operations such as breaching defenses or defending assets by intercepting projectiles \cite{spencer:2019}. Upon closer consideration, many of the challenges faced when automating these operations, such as “dirty” data, i.e. data with low signal to noise ratios, and sparse data, i.e. small training data sets~\cite{henz:2018}, are also faced when performing civilian rescue operations. These resiliency challenges will be illustrated below with a small military rescue vignette and correlated with current Army research efforts and gaps including sensor fusion, autonomous coordinated “swarms” \cite{piekarski.2016}, and resource constrained computing. \cite{im:2019}

Consider for a moment a complex future operating environment \cite{army:2019} where military operations take place in the dynamic cyber and physical environments of large urban areas. AI algorithms often require training from large amounts of data for maneuver to include a priori knowledge of infrastructure including roads, buildings, and subterranean passageways. During military operations urban environments can change rapidly as buildings are destroyed and barriers to movement are erected, leaving little existing knowledge for aiding autonomous maneuver. In addition, instability of remaining structures is likely compromised so situational understanding (SU) must be gained on-the-fly, all while an adversary is actively employing anti-access and area denial (A2AD) capabilities. It is against this backdrop that future autonomous system will be called upon to locate, extract, and maneuver to safety either human teammates or other autonomous systems. Gaining SU in this environment will require fusing data from multiple sensing modalities over communications links that are unreliable at best, or denied at worst, requiring AI that achieves consensus on courses of action with incomplete information. Maneuvering through unstable structures requires understanding of the physical world including solid body mechanics, material strengths and failure modes, and physical systems. The rescue of people or equipment is an operation that embodies these challenges for AI and autonomous systems to perform.

There are early research efforts to begin addressing many of these challenging areas including developing new sensor modalities, sensor fusion, and robotic perception. In order to complete the task illustrated above, to perform a military rescue operation, in the complex dynamic urban environment will require significant effort in many areas including the integration of these efforts into complex systems of systems. For instance, open challenges exist when sensor fusion occurs with asynchronous data, unavailable data, or active deception.  Open challenges also exist for autonomous maneuver in dynamic environments, e.g., off-road driving, and when interaction with the physical environment is required such as moving obstacles or sliding a chair to remove an obstacle. Finally, communication and coordination of cooperating sensors and autonomous systems when traditional modes of communication such as radio frequency (RF) links are unavailable or denied is a long standing challenge that is compounded for this scenario by the uncertainty of the physical environment. Advancements that address these challenges will support civilian rescue operations such as in natural disasters without endangering additional human lives of rescuers and the compression of timelines for rescue operations.

%% file: resilient-design.tex
 \section{Resilience by Design}
\label{sec:resilient-design}

\saurabh{Contributors (open to suggestions): Lead -- Milind Kulkarni; Aly El Gamal, Suman Jana, Prateek Mittal, Xiaokang Qiu, Somali Chaterji}
% Karen Marais

In this section, we discuss some design approaches to autonomous software systems that can make them resilient to the set of perturbations introduced earlier. For each we describe the problem context, some of the most promising techniques being researched today, and the desired end state and open challenges that we have to tackle to get there.
This section includes discussion of attacks against building blocks of autonomous systems, resilient ML algorithms, immune-inspired resilient algorithms, program specification for resilience and keeping in mind human-in-the-loop.  

We give in the side bar a distillation of the significant problems that we discuss in this section and the solution approaches. For the solution approaches, we categorize them as {\em Nascent} or {\em Developing}. The former means that there is little work so far but there is growing interest and some early promising results; the latter means there is a sizable body of work but it is growing and final answers are still in the future. By the nature of this article, none of the problems being discussed have fully mature solutions. 

\noindent \fcolorbox{black}{orange}{
\begin{minipage}{0.9\linewidth}
{\bf Problems:} \\
1. Attacks against ML building blocks \\
2. Ease of generation and transferability of adversarial examples \\
3. Vulnerabilities of stochastic algorithms to corner cases \\
4. Synthesis of programs for resilience \\
5. Formal verification of correctness properties
\end{minipage}
}
\definecolor{celadon}{rgb}{0.67, 0.88, 0.69}
\noindent \fcolorbox{black}{celadon}{
\begin{minipage}{0.9\linewidth}
{\bf Solution Approaches:} \\
1. Whitebox testing and verification \boxed{N}\\
2. Optimization for resilience (in addition to accuracy)  \boxed{D}\\
3. Evolutionary algorithms with resilience objectives  \boxed{N}\\
4. Scalable verification techniques  \boxed{D}\\
5. Verification with human-in-the-loop \boxed{N} \\ \\

{\em \boxed{N}: Nascent; \boxed{D}: Developing}
\end{minipage}
}

% \todoinline{Each person can write 0.5-1 page under the headings according to the text above. To make it easier, just write in one place under 2 headings: ``Current state'' and ``End state and ways to get there''.}

% \todoinline{Aly, Prateek: ML algorithms being resilient to adversarial attacks; agile in being able to handle unexpected inputs/conditions; resilient in being able to handle divergence between training and production data. 

\subsection{Attacks against building blocks of autonomous systems}
\saurabh{Material from Aly}

\noindent {\bf Problem Context}. \\
Deep learning algorithms have been shown over the past decade to be very successful in various image and speech processing applications (see e.g.,~\cite{dl}), and more recently for wireless communication systems (see e.g.,~\cite{radioml1} and \cite{radioml2}). These success stories suggest the applicability of deep neural networks in a ubiquitous fashion in the near future. However, for this to happen, such algorithms have to be designed while taking into consideration potential exposure to adversarial attacks, especially with their recently discovered vulnerabilities (see e.g., \cite{intprob}). Furthermore, these adversarial attacks are effective even when the attacker can only perturb the test data, and even there only a small part of each data point, as in evasion attacks (see e.g.,~\cite{eattacks}). One of the early and most efficient attacks that have been proposed in the literature is the Fast Gradient Sign attack (FGS)~\cite{fgs}. FGS highlights the vulnerability of neural networks, as the adversarial perturbation takes place in the direction of the gradient of the loss function:
\begin{equation}\label{fgs_eq} 
\widetilde{x} = x + \eta * {sign}\left(\nabla_x J(w,x,y)\right).
\end{equation}

In (\ref{fgs_eq}), $x$ refers to the original input sample and ${sign}\left(\nabla_x J(w,x,y)\right)$ is the sign of the gradient of the employed cost function $J(w, x, y)$, which is a function of the input, $x$, the desired output, $y$, and the classifier weights, $w$. The parameter $\eta$ is typically an $l_p$-bounded perturbation. 

More stealthy attacks, that typically incur significantly higher computational cost, than the FGS attack have been proposed. 
% \saurabh{I do not understand "yet less computationally efficient".}
Important examples are the evolutionary-algorithm-based attack introduced in~\cite{attack}, the feature-selection-based Jacobian Saliency Map attack introduced in~\cite{papernot}, and the iterative Deep Fool and Carlini-Wagner (CW) attacks introduced in~\cite{deepfool} and~\cite{cw}, respectively. In particular, iterative application of variants of the gradient sign concept were shown to present more effective attacks than the one step application of~\eqref{fgs_eq}. This issue was closely analyzed in~\cite{pgd}, where it was suggested that the one step FGS attacks leaks information about the true label. The classifier can then learn to perform very well on these adversarial examples by exploiting this leaked information. We are inspired by such analysis as it relies on characterizing the flow of label information to assess the effectiveness of an adversarial attack. More generally, we believe that a principled approach to quantifying the value of knowledge at both the attacker and defender, as well as characterizing the flow of important information regarding the output of the machine learning models and input sample distribution, is missing in the literature, and crucially needed to reach a deeper understanding of the resilience of the foundational blocks of autonomous systems.
% \saurabh{Aly can you add some material to address the remaining points from the structure above (promising current directions, open research problems). \textcolor{red}{Aly: I edited that part according to Saurabh's comment. Please let me know if more edits are needed.}} 
% \saurabh{Resolved}
% \todoinline{[Aly] Bring in the material from later in the section about open challenges and include that here. We are putting all discussion about a particular topic in one place.}
% \todoinline{Suman: resiliency by systematic testing and verification of ML, testing and verification challenges in ML-based autonomous systems, and some directions. Added a draft below. \textcolor{red}{Saurabh: DONE}}

% \textcolor{red}{Prateek: I am inserting text below on recent advances in adversarial examples, covering black-box attacks, physically-realizable attacks, and also open-world machine learning contexts. Aly -- please take a look if you get a chance, thanks! }

% \textcolor{red}{Aly: I checked it carefully, and it looks good. Just made minor edits!}

\noindent {\bf Current State}.\\ 
Recent work has also demonstrated the practical threat of adversarial examples in the context of real-world systems and constraints. For example, Papernot {\em et al.}~\cite{papernot:asiaccs17} considered the scenario of a \emph{black-box} threat model, in which the adversary does not have access to the details of the internal model structure or parameters, and also does not have access to the training data. Even in this challenging setting, the threat of adversarial examples persists. For example, it is possible for an adversary to locally train a model based on synthetically generated data, with associated labels obtained from interacting with the target model. The adversary can then use the locally trained model to generate adversarial examples, using standard techniques such as the FGS and CW attacks discussed above. This approach exploits the phenomenon of \emph{transferability} of adversarial examples: with high likelihood, adversarial examples generated using the adversary's locally trained model successfully induce mis-classification on the actual target model. Bhagoji {\em et al.}~\cite{bhagoji:eccv18} explored an alternative attack approach for the black-box setting, which does not rely on transferability. Instead, the gradient term in~\eqref{fgs_eq} is replaced by an approximate estimate of the gradient which can be computed in a black-box manner by interacting with the target model.

Another thread of research has considered the question of designing physically-realizable adversarial examples for autonomous vehicles, whose effects persist in presence of real-world environmental constraints such as varying depth of perception, varying angle of perception, and varying brightness conditions.
Standard attack techniques discussed previously produce adversarial examples that do not work well under such real-world conditions. Eykholt {\em et al.}~\cite{eykholt:cvpr18} and Sitawarin {\em et al.}~\cite{sitawarin:dls18} modify the attack optimization problem to include such varying real-world conditions: their key insight is to incorporate a set of image transformations such as perspective transformations, image resizing, and brightness adjustment - as dataset augmentation techniques - while designing adversarial examples and evaluating their efficacy.

\noindent {\bf Desired End State and Ways to Get There} We note that machine learning driven autonomous systems need to be resilient not just to perturbed variants of the training or test data, but also to \emph{unexpected} inputs, also known as \emph{out-of-distribution examples}, that do not lie close to the training or test data. After all, applications such as autonomous vehicles operate in a dynamic environment and may naturally encounter objects that were not part of the training/test data. This observation has motivated a line of research on \emph{open-world} machine learning, which augments conventional classifiers with another classifier for first deciding if the input is an in-distribution sample or an out-of-distribution sample~\cite{odin:iclr18,cabana2019multivariate}.  Approaches such as ODIN~\cite{odin:iclr18} rely on prediction confidence to make a determination about a sample being in-distribution or out-of-distribution. Detected out-of-distribution samples can simply be rejected. However, recent work by Sehwag {\em et al.}~\cite{sehwag:aisec19} has shown that existing open-world learning frameworks are not robust: an adversary can generate adversarial examples starting from out-of-distribution data to bypass detection. Here, instead of detecting whether the input sample is in-distribution or out-of-distribution, we detect whether each hidden layer, while processing the input sample, is in-distribution or out-of-distribution, as we have shown recently in~\cite{sivamani2020non}.

Recently, there has been rapid progress in discovering effective defense strategies. However, most of these defense techniques are based on custom tailoring to the employed machine learning model, and the reported robustness would not hold, not only if the model's architecture change, but even if its parameters change due to new training data. For instance, the currently proposed approaches for defending against the FGS attack as well as iterative variants like the CW attack rely on simulating the attack. Furthermore, these defense strategies assume knowledge of the attack strategy while designing the defense, and provide little or no guarantees if the attacker decides to make, even very slight, changes to its strategy (e.g., choosing another model than the target model). Finally, while there is a vast literature on combating adversarial \emph{intentional} machine learning attacks, there is little available work on designing the models to be resilient to unexpected inputs, that could result for example from unexpected behavior of other system components.
% , and could even be fed to the machine learning model while training. 

% \textcolor{red}{Aly: I moved the part below from the end of Section 4}

We believe that there are three key aspects regarding the challenge of designing resilient machine learning algorithms: {\bf 1: Perceptibility of the attack.} In the case of images, this is easily defined through human perception. For other applications, straightforward detection mechanisms should be used to judge how easy is it to detect the presence of an attack. {\bf 2: Value of knowledge in adversarial and uncertain environments.} For example, how to best exploit a finite-length secret key to protect a machine learning model. {\bf 3: Computational cost associated with attacking a target model.} Modern-day cryptography approaches rely on high computational cost associated with breaking their defense. Ideally, if the attacker has full knowledge of the system and the defense strategy, as well as an unlimited computational power, then there is no hope for resilience beyond the fundamental limits of the learning model; see e.g., \cite{nipsprateek} for the design of classifiers that consider such attacks. 
%We believe that there are two key problems regarding the challenge of designing resilient machine learning algorithms: 1- Quantifying the value of knowledge in adversarial and uncertain environments. For example, how to best exploit a finite-length secret key to protect a machine learning model, 2- Identifying the computational cost associated with attacking a target model. Ideally, if the attacker has full knowledge of the system and the defense strategy, as well as an unlimited computational power, then there is no hope for resilience. For example, modern-day cryptography approaches rely on high computational cost associated with breaking their defense. 
% SB (11/17/19): Worked till here.

\subsection{Resilient ML algorithms}
\saurabh{Material from Suman}

\noindent {\bf Current State}
Over the past few years, significant advances in ML have led to widespread adoption and deployment of ML in security- and safety-critical systems such as
self-driving cars, malware detection, and gradually in industrial control systems.
% ***KM: I'm pretty sure it hasn't been used in actual ACAS, only proposed for it. Getting ACAS changes approved is a long process***. 
% SB (9/22/19): Changed. 
However, ML systems, despite their impressive capabilities, often demonstrate unexpected/incorrect behaviors in corner cases for several reasons such as biased training data, overfitting, and
underfitting of the models. In safety- and security-critical settings, an attacker can exploit such incorrect behaviors to cause disastrous effects like a fatal collision of a self-driving car. Even
without an attacker trying to cause harm, a Tesla car in autopilot recently crashed into a trailer
because the autopilot system failed to recognize the trailer due to its “white color against a brightly lit sky.” 

Existing ML testing approaches rely mostly on manually labeled real-world
test data or unguided ad-hoc simulation to detect such corner-case errors. But these
approaches do not scale well for real-world ML systems and only cover a tiny fraction of all possible corner cases (e.g., all possible road conditions for a self-driving car). A promising and active line of research is to build a novel set of testing and verification tools for systematically finding such cases and ensuring security and safety of ML systems.

Our key insight is that most limitations of existing ML testing techniques result from their blackbox nature, i.e., they do not leverage an ML system’s internal behaviors
(e.g., outputs of intermediate layers) to guide the test input generation process. We have been developing
% ***KM: probably shouldn't be proposing research here?***
% SB (9/22/19): Changed
whitebox testing and verification tools for performing static/dynamic/symbolic analysis
of ML systems~\cite{tian2018deeptest, wang2018formal, wang2018efficient, pei2017deepxplore}. 
These types of analyses have been used successfully for testing and verification of traditional software. However, the existing tools are not suitable for testing/verifying ML
systems for the following reasons. First, traditional software logic is written by the developer while the logic of ML systems is inferred automatically from training data. Next, unlike most traditional software, ML systems tend to be highly non-linear. Finally, for some of these autonomous systems, it is a challenge to specify what is the expected behavior, considering the large number of possible interactions among the cyber, physical, and human elements. 
Therefore, support will be useful for specifying envelopes of desirable (and/or undesirable) behavior from these systems, which should then be decomposed into desirable or undesirable output states from each constitutent algorithm.
% manually
% creating specifications ***KM: haven't explained where writing specs comes in*** for complex ML systems like autonomous cars is infeasible as the underlying logic is too complex. 
% SB (9/22/19): Changed

Over the last three years, we have been building new testing and verification tools to bring more rigor to DL engineering~\cite{tian2018deeptest, wang2018formal, wang2018efficient, pei2017deepxplore, lecuyer2019certified}.
Given the challenges in specifying a full functional spec of DNNs, we design our tools to check transformation-invariant properties such as “slight light condition change must not change the image class." We have explored different design tradeoffs between scalability, completeness, and soundness.
Our tools have found thousands of corner-case errors in different DL systems including state-of-the-art image classifiers, object detectors, malware detectors, self-driving car software, and cloud computer vision systems built by Google, Amazon, IBM, and Microsoft. We have also been able to verify
some of these DL systems on popular datasets. Our testing and verification tools often outperform other existing tools by orders of magnitude (5,000x on average). We are
encouraged to see that the concepts and algorithms in our
tools have already started to gain adoption by other research groups and the
industry~\cite{odena2018tensorfuzz, zhang2018deeproad, cohen2019certified}.

\noindent {\bf Desired End State and Ways to Get There}
Despite the promising initial results
are, there are many difficult open challenges that are not yet addressed.
For example, existing DNN verification tools focus on verifying properties on a limited set of test samples with the hope that the guarantees achieved on individual samples generalize to unseen samples. One way to minimize such assumptions is to try to  adapt existing specific testing and verification techniques (e.g., interval analysis, mixed-integer programming) to reason about distributions of inputs instead of individual inputs. Another interesting direction is to support a richer set of safety properties, different types of neural networks (e.g., RNNs), and different activation functions such as Sigmoid and tanh.

\subsection{Immune-inspired resilient algorithms}
\saurabh{Material from Somali}

% \todoinline{Somali: immune-inspired algorithms (fast action plus memory inbuilt from past attacks) and evolutionary algorithms (learning over time).}
% \todoinline{[Somali] Shorten to 1 page. Motivate why immune-inspired algorithms are good for resilience. Shorten background. Add material for: promising current directions, open research problems. 

% Shortening is not my forte but I have added more ``resilience'' aspects. I will try a chopping pass later.}

Immune-inspired algorithms fall in one of the following three sub-fields: \textit{clonal selection}, \textit{negative selection}, and \textit{immune network algorithms}. These techniques are commonly used for clustering, pattern recognition, classification, optimization, and other similar ML domains.
They are relevant to design of resilient systems because they are (under domain-specified assumptions) able to adapt to uncertain system conditions or unexpected inputs. 
%talk about how they confer resilience (done)

Often, these immune-inspired algorithms start with processes reminiscent of natural selection deploying the following steps: \textbf{initialization}, \textbf{selection}, \textbf{genetic operations} (encompassing crossover and mutation), and \textbf{termination}, which can occur when the algorithm has reached a maximum runtime or a set performance threshold. Each of these steps mimics a particular phase in the natural selection process. Further, in the selection phase of immune-inspired algorithms, there is a metric such as fitness function that measures how viable the solution is. 
Relevant to our discussion, the fitness function should incorporate metrics for resilience rather than just raw performance, \eg how does the transformation make the system more immune to new kinds of attacks. 
% While creating the fitness function can be a difficult design step and may need domain knowledge, encoding domain knowledge into the fitness function, either implicitly (e.g., design of data structures, defined constraints) or explicitly (e.g., parameter control, population initialization) can actually improve real-world resilience of these solvers. This is especially relevant as the amount of labeled data available has been rising, seemingly decreasing the reliance on domain knowledge. However, often, the labeled data available is of poor quality and in these settings domain-adapted models with domain knowledge integrated in model training has been found to increase the performance of the model relative to a model trained on synthetic data. Another advantage of this process is in transfer learning where the data distribution during training and testing are different resulting in \textbf{domain-invariant learning capability}~\cite{DSN-NIPS2016}.
Alternately, reinforcement learning (RL) can enable the selection of the best fitness function and also confer resilience to the algorithms. This is especially helpful in the case of evolutionary algorithms such as genetic algorithms being used to solve difficult optimization problems where possible drawbacks are the long time to convergence and the possible convergence to a \textit{set of fitness functions} on the Pareto frontier, rather than to a single optimal point. %For example, there can be multiple fitness functions with the end result being that we end up with a \textbf{set of optimal solutions} rather than a single optimal point, also called the \textbf{Pareto frontier}. 

\noindent \textbf{Evolutionary algorithms:} Evolutionary algorithms are useful because they can be used to search for resilient operating points of computing systems in a manner that does not incorporate strong assumptions about the behavior of the underlying system. There are essentially three very similar evolutionary algorithms: genetic algorithms (GA), particle swarm optimization (PSO), and differential evolution (DE), with GA more suitable for discrete optimization, while PSO and DE being natural fits for continuous optimization processes. In general, for evolutionary algorithms, after initialization, the population is evaluated and stopping criteria are checked. If none are met, a new population is generated, and the process is repeated till the criteria are met. For increasing the resilience of these algorithms, diversity-aware variants of these algorithms have been proposed~\cite{preparing2018}. 
% Also, domain knowledge can also be encoded in the optimization process. In fact, evolutionary algorithms are increasingly being used for problems that are non-linear, discontinuous, and in high-dimensional spaces, and they are often thought of as universal optimizers. In such diverse situations, encoding domain knowledge may also result in faster convergence, especially for application to more latency-sensitive problems or where the labeled data is absent or of poor quality. Further, 
In most current applications, the optimization process occurs offline, however, some recent systems have evolved to combine offline training with further online adaptations, such as in our recent work on optimization of configuration parameters of database systems in the face of dynamic real-world workloads~\cite{rafiki-17, sophia-atc19}. Our systems combine offline training of the neural network with \textit{online adaptation using time-efficient genetic algorithms} to search for discrete optimized state spaces. Such a design makes the system more agile to real-world variability, such as intercepting dynamic, fast changing workloads querying a database.

\noindent {\bf Desired End State and Way to Get There}
For resilient system design, we would want static training and configuration of the system to be complemented with dynamic learning, even in the face of sparse data. 
Drawing inspiration from the immune system, one can think of the following characteristics for the algorithms that may confer real-world resilience: \textit{stochasticity} (increasing the exploration space); \textit{reinforcement learning}, which can contribute to emergent behavior without global coordination; \textit{stigmergy} where the ``agents" interact with the environment. Overall, these characteristics result in an emergent and probabilistic behavior, with redundancy and adaptivity in real-world settings, which are ideal for resilience.

\subsection{Program Synthesis for Resilience}
\saurabh{Material from Xiaokang}

% \todoinline{Xiaokang: program synthesis for resilience; how to convert user requirements into formal specification so that program synthesis can work on it. \textcolor{red}{Saurabh: DONE}}

A resilient software system needs to be able to rapidly adapt itself for perturbations. This is essentially a complex and intellectually demanding programming task.
Program synthesis, the process of automatically generating programs that meet the user’s intent, holds promise to automate this programming task and significantly increase the level of autonomy.
The last decade has seen tremendous progress in the efforts of program synthesis techniques, including the synthesis of SQL queries~\cite{qbs}, cache coherence protocols~\cite{TRANSIT} and network configurations~\cite{genesis,netcomplete}, or productivity software like Microsoft Excel~\cite{flashfill,flashfill2}.
% \saurabh{Can you make the connection to resilience more direct?}

A major challenge for program synthesis is how to obtain a precise specification that reflects the programmer's goals. % This problem is also referred to as the Intention pillar of machine programming and believed to be necessary to transform the programming landscape~\cite{three-pillars}.
Toward addressing this challenge, programming-by-examples (PBE) has been an appealing technique~\cite{singh2016blinkfill, singh2015predicting}. A PBE system is given a set of input-output examples and tasked to find a program whose behavior matches the given examples through iterative interactions with the user.
Another promising technique is sketch-based synthesis~\cite{program-sketching}, in which the programmer specifies a synthesis problem as a sketch or template, which is a program that contains some unknowns to be solved for and some assertions to constrain the choice of unknowns.
While the combination of PBE and sketch-based synthesis has seen many successful applications~\cite{autograder,sketch,storyboard,Chaudhuri:2014:BBQ:2535838.2535859,pasket}, these techniques do not immediately allow the programmer to describe the resilience aspects of the target program, which are usually quantitative and optimization-oriented.
For example, a critical assumption of PBE is that the user knows what the expected output is, at least for some sample inputs. In the resilience context, however, the programmer usually cannot quantitatively determine how resilient a program is. Similarly, providing a sketch of the desired resilient program is also challenging because the programmer may not know how a resilient program looks like.
Therefore, automatic programming for resilient systems requires novel, user-friendly modalities for describing resilience-related objectives.
% \saurabh{Xiaokang, can you expand on the previous two sentences to make the connection to resilience more direct?}
% \textcolor{orange}{XQ: Done.}

% \subsection{Resilient specifications in the face of humans}
% \saurabh{The two sub-sections are combined.}

% \todoinline{[Milind, Karen] The points below need to be fleshed out in narrative form. \textcolor{red}{DONE}}

% \saurabh{Points below are to be fleshed out.}

While there has been substantial effort in formally verifying hardware and software, these efforts have largely focused on functional correctness: ensuring that the system has the functionality you want. The problem is that verification is only as good as your specification: if your specification leaves details out (e.g., some functionality is unspecified), then formal guarantees do not address that aspect of your system at all. More importantly, {\em even if all the functionality of your system} is captured, specifications often do not consider {\em non-functional} aspects: timing, energy usage, interaction models, etc. These gaps in the specification leave open vulnerabilities. For example, underspecifying the timing information of the system may leave timing side channels open, allowing attackers to glean information about a system or even perturb its behavior. A specification that does not account for the interaction model of a system may not properly account for the ways that a human interacts with a system (e.g., not considering certain classes of inputs because the specification designer does not account for humans providing perverse inputs). In the context of this under-specification, no amount of verification can help provide resilience. 

Of course, as specifications get more complex to account for all of the possible vulnerabilities and interactions, the {\em scalability} of formal techniques comes under pressure, and formal verification becomes considerably harder. Indeed, this means that in the presence of difficult verification tasks such as modeling systems that involve human interaction, practitioners often use highly simplified models of the system under inspection: easing the verification task by reducing the fidelity of the verification. While this ``solves'' the scalability problem, it leaves open the question of {\em how} to do this model simplification: a model must still capture enough behavior of the system for the formal guarantees to be meaningful. Deciding how to model systems, therefore, becomes a serious bottleneck for verification, and there have correspondingly been only a handful of studies of formal verification for human-in-the-loop systems.

Nevertheless, we identify verification of human-in-the-loop systems as an important challenge for the design of future systems, especially as more and more systems represent a collaboration between automation and humans. Consider, for example, airline auto-pilot systems that expect certain input behavior from humans but fail when the system enters an unexpected regime that causes humans to apply unexpected control inputs.

\noindent {\bf Desired End State and Way to Get There}

We propose a couple of promising directions to pursue in this space. First, can models be automatically developed for human-in-the-loop systems? Is it possible to automatically simplify a complex model (that might be automatically generated from observations of a system) to target particular desired properties such that humans can then intervene only to ensure that the model is faithful? As an example, consider observing the throttle inputs to a vehicle along with observing the motion of that vehicle to automatically infer a model relating the controls to the state of the vehicle. Second, can we tackle the scalability problem by {\em detecting simpler properties?} Rather than pursuing full formal specification and verification, it may be sufficient to make human-in-the-loop systems more robust by automatically identifying and flagging unexpected behavior (e.g., a conflict between the current state of an airplane and the types of control inputs being applied by a pilot), relying on the {\em presence} of a human in the loop to perform more fine-grained corrective action?

Constructing resilient systems with human in the loop further raises new challenges. 
The system designers should formally specify the \textit{envelope} behaviors of humans and expected by humans. 
For instance, the design of a driver-assistance system may specify the maximum tolerable latency in recognizing objects to be 10 ms; 
an interactive image retrieval system may specify the minimum time that the user is given to annotate an image; 
an AI-based auto-completion code composer may specify the maximum human input rate is 50 program tokens per minute. 
With such explicit specifications of human behaviors, compiler and runtime may reason about system resilience by taking into human factors into account. 
For instance, they may assert if the human users are given enough think time to react to the detected anomaly, 
or if the human users are overwhelmed by the amount of training samples they have to annotate. 

As the demands of resilience, and formal design principles for resilience, grow, we believe a key problem that should be tackled is scalability. How can larger systems be verified? How can more complex specifications be verified? Are there modular approaches to specification such that the verification task can be tuned to the particular property that we desire to address?

%% file: execution-platforms.tex
\section{Resilience by Reaction}
\label{sec:execution-platforms}

\saurabh{Contributors (open to suggestions): Lead -- Saurabh Bagchi; Vaneet Aggarwal, Fred Douglis, Jiawei Han, Hank Hoffmann, Felix Lin, Somali Chaterji}

% \todoinline{Each person can write $\approx$1 page under headings of ``Problem and Current State'', ``End State and Ways to Get There''. Add a title to your part and put your initial. See example below for ``SB''.}

Here we talk about the online measures to deal with  perturbations to ensure that as much of the system functionality as possible is maintained. We structure our discussion in terms of the execution platforms (mobile, edge, cloud, or HPC) and the algorithms that are executing. That is, we consider what changes can be made to either the execution platform or the algorithms to ensure that the cyber system continues to operate in a resilient manner despite the occurrence of perturbations at runtime. 

Like in the design section, we  give  in  the  side  bar  a  distillation  of  the  significant problems  that  we  discuss  in  this  section  and  the  high-level solution approaches being developed.  For  the  solution  approaches,  we  categorize them as {\em Nascent} or {\em Developing}. 

\noindent \fcolorbox{black}{orange}{
\begin{minipage}{0.9\linewidth}
{\bf Problems:} \\
1. Live reconfiguration or adaptation of distributed applications \\
2. Detecting anomalies in streaming textual data \\
3. Approximate computing while meeting resilience guarantees \\
4. Handling stragglers in distributed computation \\
5. Robust policies for adapting autonomous systems
\end{minipage}
}

\definecolor{celadon}{rgb}{0.67, 0.88, 0.69}
\noindent \fcolorbox{black}{celadon}{
\begin{minipage}{0.9\linewidth}
{\bf Solution Approaches:} \\
1. Live reconfiguration of distributed applications \boxed{N}\\
2. Monitoring and optimizing network middle boxes  \boxed{D}\\
3. Stream data mining with soft real-time guarantees  \boxed{N}\\
4. Flexible approximation and execution on a variety of platforms \boxed{D}\\
5. Straggler mitigation techniques being data-centric, heterogeneous, and proactive \boxed{D} \\
6. Specification of high-level resilience goals for autonomous systems \boxed{N} \\
7. ML + Control Theory to achieve these goals \boxed{N} \\ \\

{\em \boxed{N}: Nascent; \boxed{D}: Developing}
\end{minipage}
}

\begin{comment}
\todoinline{Fred: Distributed enclave defense using Configurable edges. \saurabh{DONE}}

\todoinline{Vaneet: Straggler Management Approaches in Cloud Computing. \saurabh{Done}}

\todoinline{Felix, Hank: multiple controllers work together to achieve the end state;  system support for multiple controller control system; system support for resilience in the presence of humans. \saurabh{DONE}}

\todoinline{Somali: tuning configurations of distributed applications for resilience. \saurabh{DONE}}

\todoinline{Saurabh: approximate computation while meeting availability requirements; flexible execution on mobile, edge, cloud. \saurabh{DONE}}

\end{comment}

\subsection{Tuning Configurations of Distributed Applications for Resilience}
\saurabh{Material from Somali}

Given the rise in data being generated by different sectors, especially in IoT and automation, scalable data engines, low-latency in-memory engines (e.g., Redis), and stream analytics engines, are on the rise.
In the context of scalable data processing engines, we can consider the changing, unpredictable workload patterns as perturbations to the system, in the face of which the system will need to react. Without such reaction, the rate of servicing the requests will drop, and in some pathological cases, requests for services will be silently dropped.

Most static database configuration tuners, such as~\cite{rafiki-17, ottertune, ituned}, tend to be ``reactive'' because they use optimization techniques for changing the configuration parameters \textit{when} there is a change in the application characteristic, e.g., change in read-write ratios for database workloads. However, for global-scale, multi-tenant execution pipelines and data repositories, such as the metagenomics repository MG-RAST~\cite{federation-17,mgrast-15}, the workloads may be more dynamic and unpredictable, making it harder to ``react'' to workload changes on the fly. The goal for the reconfiguration is to maximize the system's performance, using metrics such as the database's throughput and tail-latency. For such cases, recently, predictive configuration tuners that can work with dynamic workloads have been designed, such as the NoSQL database configuration tuner \textsc{Sophia}~\cite{sophia-atc19} and \textit{Optimus Cloud}~\cite{optimus-atc20}. \textsc{Sophia} incorporates a workload predictor \textit{and} a cost-benefit analyzer (CBA) in the optimization protocol that takes into account the cost of a reconfiguration (transient dip in throughput, possible transient unavailability of data) and the benefit (improved performance due to more optimized configuration). Subsequently, the system's configuration parameters are changed only when the CBA determines that the benefit outweighs the cost of reconfiguration. Then the system implements a graceful, decentralized scheme for the reconfiguration, so that data never becomes unavailable or is availability-aware, in sync with the organization's service-level agreement (SLA).
In the context of analytics workloads running on cloud platforms, some current works, Selecta~\cite{klimovic2018selecta} and Cherrypick~\cite{alipourfard2017cherrypick} can optimize the parameters of the cloud computing environment by predicting what will be optimal for the just-arrived application. Naturally, all this is predicated on accurate enough prediction of changing workload patterns.

\noindent{\bf End State and How to Get There}

The continuing challenge remains to perform live upgrade of systems due to various events (changes in workload characteristics, data availability or consistency requirements, spatial migration of computing equipment), without degrading the data availability. An additional dimension to this problem is introduced by the use of cloud-hosted software, including some hybrid solutions, where part of the execution is on-premises and part on a remote cloud platform. The update could be to the configuration parameters, or more intrusively, to the software itself.

We can get there by learning from the long line of work on live upgrades of software systems, primarily focused on single-node software~\cite{soules2003system,hicks2005dynamic}. We will have to bring in new distributed protocols that can progressively upgrade a distributed application, while keeping data continuously available and while respecting any end-user consistency requirement (SLA). It will be important to bring in a predictive component to such work, so that the reconfiguration can be initiated prior to the event, but in anticipation of it. Such proactive action can ensure high availability as well as decision as to whether the reconfiguration is beneficial at all. With respect to cloud deployments, cloud providers have mechanisms for data migration and VM migration but they have costs in terms of performance or simply \$ costs. Therefore, the prediction can determine whether the benefit normalized by the \$ cost or the transient performance impact is tolerable to the application. 

\subsection{Distributed Enclave Defense Using Configurable Edges}
\saurabh{Material from Fred}

\noindent{\bf Problem and Current State}
In a geographically distributed system, the state of the network can change rapidly due to any of a number of factors.  As mentioned in the introduction, perturbations can result from failures, overt attacks, or simply competition for resources.  In most environments, there is  no central arbiter with global knowledge of  the state of the network.  Instead, endpoints at the edges must infer network characteristics and adapt to changes to those characteristics.  This adaptation can take many forms, including rerouting traffic, transforming content, and others.

There is a long history of adaptation on a case-by-case basis.  For instance, Fox et al. described a mechanism for dynamic transcoding of web images into low-resolution forms~\cite{fox1996adapting}. Split TCP~\cite{bakre1995tcp,balakrishnan1997comparison} separates a TCP connection into one connection between a client and a proxy, and another connection between the proxy and a server; this allows the proxy to treat each part of the connection appropriately for its characteristics, such as high delay or loss.  Middleboxes~\cite{sekar2011middlebox} are a generalization of this approach, interposing for various optimizations including Split TCP~\cite{le2015experiences}.  
% \saurabh{Would a more recognizable optimization than Split TCP be apt here?}
% \fred{I think that is a pretty widely-known technique, but maybe I'm biased.  Alternatives welcome, but nothing comes to mind for me.  Also: is this the only thing I need to respond to at this point?  Not that I'm complaining if so...}

A holistic view that deals with dynamic changes to the network topology and workloads is more challenging.  The DEDUCE\footnote{DEDUCE stands for \emph{Distributed Enclave Defense Using Configurable Edges}.} system from Perspecta Labs adds a ``bump in the wire" between edges of a network and the WAN.  The DEDUCE box is a middlebox that monitors system behavior and performs a number of optimizations.  It splits connections for TCP and UDP , allowing it to transparently reroute via other DEDUCE edge nodes, change characteristics such as TCP congestion optimizations (e.g., from Cubic~~\cite{ha2008cubic} to BBR~\cite{Cardwell:2017:BCC:3042068.3009824}, forward error correction, transcoding, and others.  To do this, it needs a strong view of the state of the network~\cite{duffield2003simple}, as well as the ability to evaluate ongoing network utility.  It continually updates its plans~\cite{chen2018radmax} for how best to achieve its goals, which may be implicit (competing best-effort flows) or explicit (information about specific flows with deadlines).  

Note that some aspects of its optimizations, such as rerouting, are similar to the functionality of the underlying IP network: IP can dynamically detect network outages and find new routes.  The difference is that DEDUCE can apply a more holistic view, for instance rerouting one flow, consistently, along a particular path and a different flow along an alternate path.  IP, by comparison, would intermix the packets along each path, leading to packet reordering and other performance implications.

\noindent{\bf End State and How to Get There}
DEDUCE is an example of a set of cooperating middleboxes that operate in isolation from the rest of the network.  That is, each DEDUCE endpoint can work with other DEDUCE endpoints to optimize traffic, but any traffic to non-participating edge networks is untreated.  For the Internet to be truly resilient, techniques such as these should be more broadly adopted.  This means for example that any communicating parties would be able to change their TCP congestion treatment dynamically (e.g., learning the best algorithm for a given situation~\cite{sivaraman2014experimental}), add or remove forward error correction or other content-level treatments~\cite{liu2002tcp}, adapt content to reduce bandwidth requirements~\cite{chandra2000application}, etc.  
%\saurabh{Cite and mention some other works in this space to make the text sound more neutral than coming from the Deduce authors}

The ability to adapt traffic is only half the solution, however.  The bigger challenge is in deciding \textit{how} to adapt, in the presence of incomplete, distributed information.  Network tomography~\cite{duffield2003simple} is still in relatively early stages, but the ability to gather and process dynamically changing state is crucial to network resilience.  A crucial aspect of this processing is being able to distinguish between failures due to resource contention and those due to hardware outages.  For instance, if losses are due to congestion, adding extra redundancy via FEC simply contributes to the congestion; if losses are due to a faulty router, redundancy may provide appropriate resilience.  Similarly, the reaction to a denial of service attack may be different from the reaction to normal high traffic.  

\subsection{Detecting Anomalies in Real Time through Text Mining}
\saurabh{Material from Jiawei}
% Detecting anomalies through text mining, in real-time.

In many scenarios, anomalies can also be uncovered through mining textual information (e.g., intruders may respond to system requests with na\"ive or threatening languages, a system may generate warning messages when detecting unusual situations, or people who observe something abnormal may signal alarms).  It is thus critical to detect anomalies through text mining, in real-time.  Preprocessing should be conducted beforehand by learning the text embedding space in typical situations with the distributions of phrases, topics, sentences, language features, and sentiments computed and stored, by using advanced text embedding methods developed recently, such as Word2Vec \cite{Mikolov+13}, Elmo \cite{Peters+18}, BERT \cite{Delvin+19}, and JoSE \cite{meng2019spherical}.  % Stream data mining methods should be conducted in real-time, and online distributions of text elements will be computed and monitored closely.  

\noindent{\bf End State and How to Get There}. There is a need to develop stream data mining methods that can operate in real-time, e.g., they can calculate in an online manner, distribution of text elements and monitor the distribution closely. 
The language features (e.g., phrases, aspects, sentences, paragraphs, or topics) that substantially deviate from the typical ones can be considered as semantic outliers and should be detected and analyzed promptly by integration of text embedding and outlier detection analysis methods \cite{Zhuang+17}.   Alternatively, one may also use classification methods to train the system beforehand by collecting text messages in previously happened abnormal situations and go against the text happening in usual situations and such trained models can be used to signal anomalies on the fly.  One can also integrate text classification and text outlier detection mechanisms to further enhance the quality of online anomaly detection.
% \saurabh{[Jiawei] Some more text about current solution approaches will be good.}

% \todoinline{Saurabh: approximate computation depending on capabilities of platform without degrading accuracy too much; do this in context-specific manner}
\subsection{Approximate Computation with Resilience Guarantees}
\saurabh{Material from Saurabh}

\noindent{\bf Problem and Current State}
Many computations are inherently approximate---they trade off quality of results for lower execution time
or lower energy. Approximate computing has recently emerged as an area that exposes additional sources
of approximation at the computer system level, e.g., in programming languages, compilers, runtime
systems, operating systems, and hardware architectures, thereby enabling us to re-define how we think
about programs that implement novel solutions to an important class of problems. This has important implications
for resilience because many demanding applications (such as, streaming video analytics)cannot
run on resource-constrained devices (such as, IoT devices) because they exhaust the limited resources
(memory, memory bandwidth, compute, IO, etc.). Therefore approximation has emerged as a potential
technology, thus broadeing the domain of possible execution platforms for a wide variety of applications.
One challenge of the
area of approximate computing has been that the accuracy and performance of applying approximate
system-level techniques to a specific application and input sets are hard to predict and control.
Today this leads to too conservative choices for approximation~\cite{laurenzano2016input},
unacceptable quality outputs~\cite{samadi2013sage, baek2010green}, and even incorrect executions~\cite{ringenburg2015monitoring}.
While the current approximate computing approaches show that the techniques have a lot of promise,
making robust predictions about accuracy and performance is a key challenge to successful adoption of
approximate computing in real-world applications.

The relevant current works in this space (approximation with resilience guarantees) answer the following
three broad questions, in one or more of the domains of mobile applications, streaming video analytics,
image processing, visualization of scientific computation, etc.

\begin{enumerate}

\item {\em When to approximate}. It searches for the period of the application’s execution that is most productive
to approximate. This is driven by early evidence that depending on when a specific technique is applied,
there may be wide variations in time to convergence of the application or the quality of the output~\cite{mitra2017phase}.
The granularity of the decision will be application specific and also subject to execution time constraints.
Finer-grained monitoring and control are likely to lead to better performance-quality tradeoff, but with a
law of diminishing returns. However, such monitoring and control come with their own overhead as well.

\item {\em How to approximate}. Any approximation technique typically accommodates one or more configuration
settings, which captures how aggressively the approximation is done. For example, with a Neural
Network-based video analytics query processing, we have to determine what is the optimal number of
layers or the level of downsampling of the video frame. As another example, for a demanding computational genomics application, we have to decide how to segment the data and process in parallel for creating an SVM model in a distributed manner~\cite{ghoshal2015ensemble}. Consider that many applications in our target
domains comprise pipeline of multiple software components or methods, each of which can benefit from
one of several approximation techniques, and each technique comes with its configuration setting.

\item {\em How to approximate in input-aware manner}. It appears from some early evidence~\cite{xu2018videochef, laurenzano2016input} that in some
important cases, the above two decisions have to be made in an input-aware manner. For example, if
one is approximating a video stream and the stream consists of relatively static scenes, more aggressive
approximation can be applied than if it is a sports scene with fast movements. Particularly, since we want
to bound the accuracy loss with approximation, it is important to take the input dependence into account.
For our target domains,
the characteristic of the input may change within the stream, requiring that the
{\em when} and {\em how} decisions be revisited.

\end{enumerate}

\noindent{\bf End State and How to Get There}
There is the need to provide sound and practical techniques to approximate computation,
under varied and unseen input data, while bounding the loss in accuracy
and providing robust estimates for reduction in energy consumption and other resources.
This will enable the grander vision off applications that can ``flit'' effortlessly between
multiple execution platforms depending on the three axes of what is the capability of
the platform, what is the resource demand of the computation (including
approximation), and what is the cost of moving computation or data and orchestrating
possibly distributed execution among the platforms.

There is the need to develop core algorithms to predict the impact of approximate computing on the accuracy
and output quality. Further, we should create the models such that they take into account the input dataset
and the state of the execution in deciding on the appropriate approximation configuration. The current
approximate computing techniques are often inflexible and may miss profitable approximation opportunities
or may mispredict the error rate. They are also fragile in the sense that their performance can fluctuate
unacceptably under different input datasets. Fine-grained input-aware approximation algorithms that
the community is developing has the potential to overcome these key challenges of approximate computing. Moreover,
there is the need to show how to combine system-level and application-specific approximation techniques in
the various target domains.

\subsection{Resilience to Stochastic Task Sizes in Distributed Computation}
\saurabh{Material from Vaneet}

Large scale computing jobs require multi-stage computation, 
where computation per stage is performed in parallel over a large number of servers. 
The execution time of a task on a machine has stochastic variations due to many contributing factors such as co-hosting, virtualization, hardware and network variations~\cite{Cheng2014IMC,xiang2016joint}. 
A slow server can delay the onset of next stage computation, 
and we call it a \emph{straggling} server. 
One of the key challenges in cloud computing is the problem of straggling servers, 
which can significantly increase the job completion time~\cite{Garraghan2016TSC, Ouyang2016SCC, Guo2017TPDS}. 
Resilience to straggling servers is essential to counter the possibility of missing deadlines in job execution. This is relevant in autonomous systems of the kinds introduced earlier because there are (soft) timing requirements and many workloads execute on cloud computing platforms. 

{\bf Current State:} %It has been observed that task execution times have significant variability, 
%partly due to resource sharing by multiple jobs~\cite{dean2013tail}.  
%The slowest tasks that determine the job execution time are known as ``stragglers''.  
Some of the key approaches to mitigate the effect of stragglers are to
have speculative execution which acts after the tasks have already slowed down \cite{Dean2008ACM} or proactive approaches that launch redundant copies of a task in the hope that at least one of them will finish in a timely manner~\cite{Anantha2013NSDI,BaditaRep}. When redundant tasks are launched on different servers, one approach is to perform an erasure-coding that provides significantly more flexibility as compared to replication. By having coding-theory based redundancy approaches, the user waits for any $k$ out of the $n$ servers to finish, while each server runs a smaller fragment of the task \cite{Aktas2018Sigmetrics,Badita2020Inf}. Coding-theoretic techniques have been proposed to mitigate the effect of stragglers in gradient computation \cite{tandon2017gradient,ye2018communication,sasi2019straggler}. %In \cite{ye2018communication}, coding techniques to reduce the running time of distributed learning tasks have been provided. Multi-stage coding techniques that uses delayed start of jobs have been considered in \cite{sasi2019straggler}.
In \cite{raviv2017gradient}, an approximate variant of the gradient coding problem is introduced, in which approximate gradient computation is done instead of the exact computation. A stochastic block code and an efficient decoding method for approximate
gradient recovery are provided in  \cite{charles2018gradient}. %A distributed computing scheme called Batched Coupon’s Collector to mitigate the effect of stragglers in gradient methods is proposed in \cite{li2018near}. 
%In \cite{halbawi2018improving}, a straggler mitigation scheme that facilitates the implementation of distributed gradient descent
%in a computing cluster is presented. They also  proposed a theoretical delay model which allows to minimize the expected running time. 

{\bf Desired End State and Ways to Get There: } Even though different approaches for straggler mitigation have been provided, efficient approaches require a holistic framework to understand the different design tradeoffs, including the completion time of the jobs, and the additional server costs spent for the jobs that will eventually not be completed. In order to come up with such a holistic framework, it would be essential to develop data-centric proactive approaches that leverage coding-theoretic and queuing-theoretic techniques. We note that deep reinforcement learning based approaches have been considered for scheduling jobs on the servers \cite{peng2018optimus,cui2017reinforcement,balla2018reliability,liu2016dependency,agarwal2019reinforcement}. Reinforcement learning approaches with speculative execution to mitigate stragglers have been considered in \cite{naik2018improving}. However, such approaches do not consider multiple jobs, heterogeneous servers,  coding-theoretic flexibilities, and approximate computing. Further, the allocation among different users must satisfy joint objectives, e.g., fairness, thereby needing decentralized, scalable solutions rather than centralized approaches \cite{agarwal2019reinforcement}. Accounting for all these degrees of freedom significantly enlarges the design space and efficient approaches that explore the design space is an interesting problem.

%These will help determine when and where to schedule each redundant task, and what task must be performed at the server. We also note that the newly launched servers can take into account the tasks for the job that have been completed so far, when the coding-theoretic flexibility is used. 

%{\bf Ways to Get There: } We require a data science based pro-active approaches that takes into account efficient coding-theoretic and queuing-theoretic approaches.   

\subsection{Adaptability with Resilience Guarantees}
\saurabh{Material from Hank and Felix}

\noindent{\bf Current State}

Computing systems must function effectively in dynamic environments where application workloads, available resources, and user requirements can all fluctuate in unpredictable ways.  To handle these dynamics, system developers create \emph{mechanisms} that enable the system to detect changes and then react to those changes.   Unfortunately, the \emph{policies} that govern how these mechanisms are applied are often ad hoc and heuristic based. These heuristics are developed by experts and tend to work very well on the system for which they were designed, but they are not robust to changes.  

As an example,
% NOTE: whoever added  this included a different type of apostrophe in the next line, which latex didn't render.  I was pretty confused for a while -- Fred
Samsung's scheduler for the Galaxy S9 smartphone has many
heuristics governing when to change clockspeed and when to migrate
a process between its fast, high-power cores and its slower, energy-efficient cores. The S9+ was anticipated
to be an upgrade with higher performance and longer battery life
due to improved processor design. However, product reviewers found that 
in practice, the S9+ exhibited lower performance and shorter life
as scheduling heuristics tuned for the S9 produced poor results on
the S9+ \cite{S9+}. The S9/S9+ is one example demonstrating how fragile
heuristic-based resource management can be, on complex,
modern processor designs, and it demonstrates the need for more
principled resource management. 

%HH: Current state is usually ad hoc and heuristic based.  These often work well for one platform, but fail when the underlying system changes in subtle ways.  Example: Galaxy S9 and S9+ OS heuristics for balancing energy and performance.

\noindent {\bf Desired End State}

Our goal is the design and development of a set of robust policies that govern how autonomous systems should react to unforeseen circumstances.  These policies should be based on well-founded principles and come with clearly stated assumptions about the conditions under which they would be expected to work and the mechanisms available for the policies to be enforced. Furthermore, we advocate for autonomous systems that do not just react, but react to accomplish some high-level, user-defined goal.  For example, such goals might be meeting a certain latency constraint with minimal energy or finding the most accurate model on an energy budget. 

\noindent {\bf Ways to Get There}
A first step to achieve the vision of adapting to high-level goals is to make those goals explicit in the program.  In other words, there should be a (possibly domain-specific) language for describing a program's quantifiable behavior and the desired range of behavior that represents successful deployment.  Furthermore, this language should also specify which behavior is a constraint---which must be respected for correct operation---and which is an objective---to be minimized or maximized subject to the constraints. In addition, this language should describe what system components can be changed to affect the goals.  The idea of specifying high-level goals and mutable program components was a key part of the Self-aware computing project \cite{self-aware}, language support for making this specification a first-class object appeared in the later Proteus project \cite{Proteus}, 
and VStore~\cite{xu2019vstore}, a video data store that respects encoding/decoding throughput constraint while maximizing storage efficiency.

Clearly defining goals is key to resilience by reaction, as it is only through the definition of goals that the system can observe they are not being met and react to restore correct operation.  Of course, there is still the question of how to react, or how to map measurements of goals into appropriate settings for the mutable system components.  We advocate a mixture of machine learning and control theory to achieve this mapping.  Machine learning models are well-suited to capturing the complex tradeoff spaces that can arise in computing systems with competing goals and many mutable components.  Control theory is well-suited to ensuring that constraints are met.  Recent work demonstrates how the two can be combined to achieve the best of both approaches \cite{caloree}.

While recent research demonstrates that it is possible to build adaptive systems to meet goals, one major challenge is unaddressed: How can multiple, independent adaptive systems collaborate effectively, if developed independently by different stakeholders?  For example, a mobile application might have goals in terms of responsiveness and image quality, while a mobile operating system might have goals in responsiveness and energy efficiency.  If those systems are developed independently, they could easily make conflicting decisions, negating their potential benefits \cite{jouleguard}.  
Thus, a common interface is likely necessary for specifying adaptive components and the goals they effect; 
a high-level negotiation mechanism are needed for coordinating their adaptations amongst different such components.

%% file: road-ahead.tex
\section{The Road Ahead}
\label{sec:road-ahead}

Here we look at the road ahead with a summary of the short-term and mid-term research and transition challenges. 

% \saurabh{Bulleted form of the points. Please add if you have ideas. I will then flesh out the bullets. The idea is to give high level directions with some example problems rather than being prescriptive about specific problems.}

\begin{enumerate}
% \addtolength{\itemindent}{-1em}
\addtolength{\listparindent}{-3em}
    \item {\em Creating resilient systems out of individually vulnerable components.} 
    There will be increasing needs to build resilient systems out of components that are {\em not} individually resilient to the perturbations that the system will have to face. Potentially there will be a large number of such components composing the system. These components will be vulnerable due to innate design and implementation vulnerabilities, or due to unpredictable interactions with the external environment, either cyber or physical. 
    
    \item {\em Speeding up the cycle of design and generation of attacks and defenses against ML algorithms.} There will be a two-pronged need in this space of resilient ML---designing algorithms that are resilient by design to a well-quantified set of perturbations and speeding up the discovery of vulnerabilities in realized implementations of the ML algorithms. The speeding up will imply a partially automated process for discovery and patching of vulnerabilities in the ML applications, as was envisioned by the DARPA Cyber Grand Challenge competition~\cite{darpa-cyber-challenge-16}.
    
    \item {\em Synthesis of resilient programs by automated means.} There is a growing body of work on automatic synthesis of programs from specifications. We have to consider that the synthesized program meets well-quantified resilience guarantees. The automatic synthesis can take the form of full program synthesis or, what is more likely, augmentation of an existing program for the purpose of increasing its resilience. In this approach, we can take inspiration from non-traditional sources, such as, immune systems in biological organisms. 
    
    \item {\em Automated configuration of increasingly complex systems, for performance as well as for resilience.} This includes efforts to automatically navigate the large space of configuration parameters and determine the close-to-optimal settings within a reasonable time bound, perhaps even online. While there is a growing body of work on automated configuration for performance, it will become important to perform such reconfiguration while meeting resilience goals, such as, server uptime and data availability.
    
    \item {\em Use of compute power close to the client devices to increase the resilience of large-scale multi-tier systems.} This involves the use of edge computing resources for redundant execution, in addition to its traditional use for reducing latency of short-running queries. We will move to some autonomous systems whose algorithms can execute in parts in each of the three tiers of execution---client devices, edge computing devices, and cloud computing devices. The partitioning can happen flexibly, even at runtime, and can be done not just for performance but also for resilience. Different degrees of redundancy will be employed for different applications and at different tiers of the hierarchy. 
    
    \item {\em Expanding the scope of approximate computation and distributed computation to include resilience as a first-order principle.} Approximate computation will become an increasingly powerful means to execute demanding applications on resource-constrained devices, or where energy resource is at a premium. We need methods to approximate computation while still being able to provide resilience guarantees, likely probabilistic. The same applies to distributed computation, which will become increasingly relevant to scale up demanding ML applications. In such cases also, the loss in accuracy relative to the centralized computation must be bounded and quantified.
    
\end{enumerate}